\documentclass[aps,prb,twocolumn,showpacs,superscriptaddress]{revtex4-1}
\usepackage{graphicx}
\usepackage{multirow}
\usepackage{color}
\usepackage{amsmath,amsfonts,amssymb,amsxtra}
\usepackage{units}

\newcommand{\tn}{$T_{\text{N}}$}
\newcommand{\ep}{$\varepsilon$}
\newcommand{\grad}{$^{\circ}$}
\newcommand{\bco}{Bi$_{2}$CuO$_4$}

\begin{document}

\title{Magnetically Induced Ferroelectricity in \bco\ }

\author{L. Zhao}
\author{H. Guo}
\affiliation{Max-Planck Institute for Chemical Physics of Solids,
N\"{o}thnitzer str. 40, D-01159 Dresden, Germany}
\author{W. Schmidt}
\affiliation{Institut Laue-Langevin (ILL), 6 Rue Jules Horowitz,
F-38043 Grenoble, France} \affiliation{J\"{u}lich Centre for Neutron
Science JCNS, Forschungszentrum J\"{u}lich GmbH, Outstation at ILL,
CS 20156, 71 avenue de Martyrs, 38042 Grenoble, France}
\author{K. Nemkovski}
\affiliation{J\"{u}lich Centre for Neutron Science JCNS at Heinz
Maier-Leibnitz Zentrum (MLZ), Forschungszentrum J\"{u}lich GmbH,
Lichtenbergstra�e 1, 85748 Garching, Germany}
\author{M. Mostovoy}
\affiliation{Zernike Institute for Advanced Materials, University of
Groningen, Nijenborgh 4, Groningen 9747 AG, The Netherlands}
\author{A. C. Komarek}
\email[]{Komarek@cpfs.mpg.de} \affiliation{Max-Planck Institute for
Chemical Physics of Solids, N\"{o}thnitzer str. 40, D-01159 Dresden,
Germany}

\date{\today}

\begin{abstract}
The tetragonal copper oxide \bco\ has an unusual crystal structure
with a three-dimensional network of  well separated CuO$_4$
plaquettes.
This material was recently predicted to host electronic excitations
with an unconventional spectrum and the spin structure of its
magnetically ordered state appearing at \tn $\sim$ 43~K  remains
controversial.
Here we present the results of detailed studies of specific heat,
magnetic and dielectric properties of  \bco\  single crystals grown
by the floating zone technique, combined with the polarized neutron
scattering and high-resolution X-ray measurements.
Our polarized neutron scattering data show Cu spins are parallel to
the $ab$ plane.
Below the onset of the long range antiferromagnetic ordering we
observe an electric polarization induced by an applied  magnetic
field, which indicates inversion symmetry breaking by the ordered
state of Cu spins.
For the magnetic field applied perpendicular to the tetragonal axis,
the spin-induced ferroelectricity  is explained in terms of the
linear magnetoelectric effect that occurs in a metastable magnetic
state.
A relatively small electric polarization induced by the field
parallel to the tetragonal axis may indicate a more complex magnetic
ordering in \bco.
\end{abstract}


\maketitle

\section{Introduction}

\begin{figure}
\includegraphics[width=1\columnwidth]{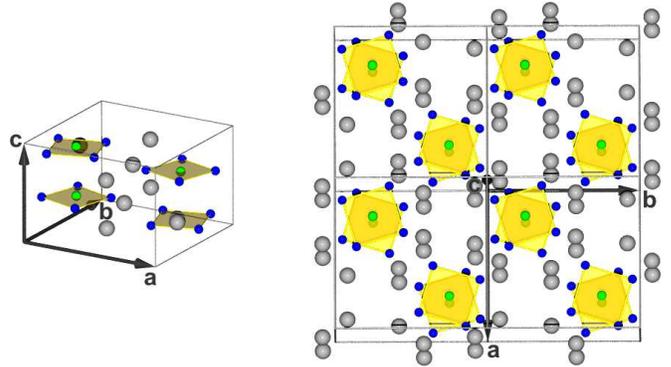}
\caption{(Color online) Crystal structure of \bco; green/blue/grey
spheres denote Cu/O/Bi atoms. \label{figA}}
\end{figure}

The discovery of high-temperature superconducting (HTSC) cuprates
triggered an extensive research on copper-based materials,
especially, low-dimensional cuprates with novel physical properties
\cite{a}. Among these materials, \bco\  is of special interest: its
``2-1-4" chemical formula is common to the prototypical HTSC
\emph{R}$_2$CuO$_{4+\delta}$ compounds  (\emph{R}~=~rare earth
etc.). However, the isovalent substitution of the rare earth
$R^{3+}$ ions by  Bi$^{3+}$ ions of similar ionic size results in a
different and unique crystal structure shown in  Fig.~\ref{figA}.
This difference might be caused by a strong covalent Bi-O bonding
\cite{b}. \bco\ has a tetragonal structure (space group
\emph{P4/ncc}), in which the Cu$^{2+}$ ions exhibit a square-planar
coordination by four oxygen ions. While in other cuprates the
CuO$_4$ units form infinite two-dimensional CuO$_2$ layers (e.g., in
\emph{R}$_2$CuO$_{4+\delta}$ \cite{a}) or quasi-one-dimensional
ribbons or ladders, as in LiCu$_2$O$_2$ \cite{c} or
Sr$_3$Ca$_{11}$Cu$_{24}$O$_{41}$ \cite{d}, the CuO$_4$ plaquettes in
\bco\ are isolated from each other. As shown in Fig.~\ref{figA}, the
CuO$_4$ plaquettes are staggered along the $c$ direction. Each
plaquette is twisted with respect to the adjacent one with a
twist/rotation angle of $\sim$33\grad, thus, forming chains in the
$c$-direction. These chains are separated by non-magnetic Bi$^{3+}$
ions.  The Cu-Cu distance along the chains is of the order of 3~\AA,
which is larger than in copper metal. Nuclear quadrupole resonance
(NQR) experiments \cite{e} indicate that the Bi$^{3+}$ ions are
involved in the superexchange between Cu ions, i.e.
Cu$^{2+}$-O$^{2-}$-Bi$^{3+}$-O$^{2-}$-Cu$^{2+}$ paths yield long
range antiferromagnetic (AFM) ordering below the N\'{e}el
temperature \tn$~\sim$~43~K.

First studies of the magnetic properties of \bco\ suggested a
quasi-1D spin-1/2 chain model with possible spin dimerizaton
\cite{f}. However, neutron diffraction measurements \cite{g,h,i,j,k}
revealed a long range three-dimensional (3D) AFM order  of C-type:
in the magnetic unit cell, which coincides with the chemical one,
spins in chains are parallel, while spins in neighboring chains are
antiparallel.  However, the orientation of Cu spins was not
unambiguously determined and two magnetic structures were proposed
-- one with spins parallel to the $c$ axis \cite{g,j} and one with
spins within the $ab$ plane \cite{k}. Optical spectroscopy
experiments including polarized far-infrared and Raman spectra on
single crystals suggest the $c$ axis orientation of copper spins
\cite{l,m,n,o}. Specific heat and thermal expansion studies of the
critical behavior around \tn\ are consistent with the 3D Ising
universality class, thus, also suggesting an easy $c$ axis
anisotropy \cite{p}. However, other experimental studies, including
the high field anisotropic antiferromagnetic resonance (AFMR)
\cite{q,r,s} and torque magnetometry measurements \cite{t} support
an easy plane anisotropy in \bco. Theoretical band structure
calculations provided conflicting predictions \cite{u,v}. Despite
many efforts, this issue remains controversial.

Spin excitations in \bco\  have been investigated  by inelastic
neutron scattering \cite{w,x}. The magnon dispersion spectrum can be
fitted using the spin wave theory with up to four AFM exchange
terms. In Ref.~[\onlinecite{x}] both intrachain and interchain
exchange interactions were found to be antiferromagnetic, which
indicates magnetic frustration in \bco.

Previous analysis of the photoelectron spectra shows that \bco\ is a
charge-transfer insulator with a large band gap (around 2~eV)
\cite{y}. The high resistivity prevents a transport study on \bco\
at low temperature. Dc conductivity measurements are only feasible
at temperatures far above \tn\ \cite{z}. Hence, dielectric or ac
conductivity measurements seem to be better suited to study \bco.
The temperature and frequency dependent dielectric response may
yield further information of microscopic physical properties of
\bco. However, so far, there exists only one report of the
dielectric properties of \bco\ in the temperature range from 100~K
to 300~K which is well above \tn\ \cite{aa}.

The discovery of giant magnetoelectric effects in multiferroic
manganites with coupled magnetic and electric dipoles \cite{ab,ac},
attracted a lot of interest in multiferroic and magnetoelectric
materials.  The spontaneous electric polarization observed in these
frustrated magnets is induced by complex spiral spin structures
which break time reversal and inversion symmetries. These unusual
magnetic states are highly susceptible to applied magnetic fields,
allowing for the magnetic control of electric polarization and, in
some cases, for the electric control of magnetization, which is of
great importance both for fundamental physics and technological
applications \cite{ad,ae}.  A similar control can be achieved in
materials showing a linear magnetoelectric effect, such as
Cr$_2$O$_3$, where an electric polarization is induced by an applied
magnetic field and {\em vice versa} \cite{xa,xb,xc}. The linear
magnetoelectric effect also requires simultaneous breaking of
time-reversal and inversion symmetry,  but is usually observed in
magnets with k~=~0 magnetic states.

Copper oxides provide a rich playground for studies of
magnetoelectric phenomena. The ferroelectricity induced by spin
spirals was observed in the multiferroic Cu-chain compounds,
LiCu$_2$O$_2$ and  LiCuVO$_4$ \cite{al,xd,xe} as well as in the
three-dimensional CuO with a high spin ordering temperature
\cite{xf}. Magnetically-induced ferroelectricity and the linear
magnetic response were recently observed in the spiral and skyrmion
phases of the chiral cubic magnet  Cu$_2$OSeO$_3$ \cite{xg,xh}.

Here, we systematically investigate anisotropic physical properties
of floating-zone grown \bco\ single crystals including magnetic
susceptibility, magnetization, specific heat, polarized neutron,
dielectric constant and pyroelectric measurements. Although the
k~=~0 AFM spin structure in \bco\ has been established decades ago,
it was not noticed before that it breaks inversion symmetry and thus
can give rise to magnetoelectric phenomena. We report the
observation of the field-induced ferroelectricity  in \bco, which we
associate with the linear magnetoelectric effect in a metastable
state. Moreover, we

\section{Experimental}

The growth of \bco\  single crystals was carried out in an optical
floating zone furnace equipped with four elliptical mirrors
(\emph{Crystal Systems Corp.}). The polycrystalline rods were
prepared from appropriate mixtures of Bi$_2$O$_3$ and CuO with high
purities. The mixed starting materials were sintered in air at
725\grad C for 72~hours with several intermediate grindings.
Finally, the obtained powder was pressed into a rod with $\sim$7~mm
in diameter and $\sim$12~cm in length under a hydrostatic pressure
of $\sim$60~MPa. Afterwards, the rod was sintered at 750\grad C for
24~hours. During the floating zone growth the crystal was grown in a
pure oxygen atmosphere. Due to the low surface tension, the molten
zone of the \bco\ is difficult to control. This might be the reason
why in earlier studies a fast growth rate was reported. Such a high
growth rate yields products that easily crack and the quality of
these crystals is not good. Therefore, we lowered the low growth
speed to 3~mm/hour while both feed and seed rods were
counter-rotated at the rates of 20-40~rpm.  Finally, we obtained
boules with typically 5~cm length. The corresponding growth
direction changed to [1~0~2] compared to the [0~0~1]-direction in
single crystals grown with faster growth rate. Our as-grown single
crystals were oriented by Laue diffraction, and cut into different
rectangular plates of 1~mm  thickness and with edges parallel to the
crystallographic principal axes.

The magnetic properties of our samples were studied using a SQUID
magnetometer (\emph{MPMS-5XL, Quantum Design Inc.}). The
measurements of specific heat were carried out using a standard
thermal relaxation calorimetric method in a commercial Physical
Property Measurement System (\emph{PPMS, Quantum Design Inc.}).

Polarized neutron scattering experiments were performed at the IN12
triple-axis spectrometer at Institut Laue-Langevin and diffuse
neutron scattering spectrometer DNS at the Heinz Maier-Leibnitz
Zentrum \cite{dnsA,dnsB}. The measurements at IN12 have been carried
out using CryoPad setup for spherical neutron polarimetry. The
flipping ratio was 21. The measurements at DNS we have done with two
different setups. In the standard setup for xyz-polarization
analysis we have studied the scattering in spin-flip and
non-spin-flip channels for three different neutron polarizations. In
this case the magnetic field at the sample was limited to the
instrument guide field (around 10~G). The flipping ratio was above
26. Additionally, we have measured the sample mounted between two
permanent magnets providing the field the sample roughly around
2.5~kG. In this case due to the depolarization of the neutron beam
by magnets, we performed no polarization analysis. The incoming
wavelength for all measurements at DNS was $\lambda$~$=$~4.2\AA.

Temperature dependent powder x-ray diffraction measurements have
been performed on a powder x-ray diffractomter (\emph{Bruker D8
Discover A25}) which is equipped with a Johansson monochromator for
Cu-K$\alpha$ radiation and which was optimized for low background
and high resolution. The measurements of our (crushed) single
crystals confirm the absence of impurity phases. The temperature
dependence of the lattice constants was measured using a He cryostat
(\emph{Oxford Phenix}).

For dielectric measurements, the plate-shaped samples were further
polished thinner to a thickness of 0.1-0.3~mm. We applied silver
paint to both sides as electrodes to form a
parallel-plate-capacitor. The samples were glued on the cryogenic
stage of  the customized probe that was inserted into a cryostat. A
high-precision capacitance bridge (\emph{AH2700a, Andeen-Hagerling
Inc.}) or a LCR meter (\emph{E4980AL Keysight Technologies}) was
used fore the dielectric measurements. The main sources of errors
such as residual impedance in the whole circuit have been carefully
compensated. In our measurements, various excitation levels (from
100~meV to 10~V) were used and no apparent difference was found,
thus, confirming the intrinsic nature of our observations.

To verify the ferroelectric nature of the possible multiferroic
transitions, the spontaneous electric polarization (\emph{P}) has
been measured via the integration of the corresponding pyroelectric
current. Firstly, we fully polarized the specimens with a static
electric field of 1000~kV/m during the cooling process from
temperatures above \tn, then removed the electric field and
short-circuited the sample at lowest temperature in order to remove
any possible trapped charge carriers. The pyroelectric current was,
then, measured during the warming process at different heating rates
(2-4~K/min).

\section{Experimental Results}

\subsection{Magnetic properties}

We measured the temperature dependence of the magnetic
susceptibility $\chi$ of a \bco\  single crystal in different
magnetic fields and field-directions (from 0.1~T to 5~T and with H
parallel to $c$- or $a$- axis). In the zero-field-cooling (ZFC) and
the field-cooling (FC) $\chi$(T) curves no essential difference was
observable.

\begin{figure}
\includegraphics[width=1\columnwidth]{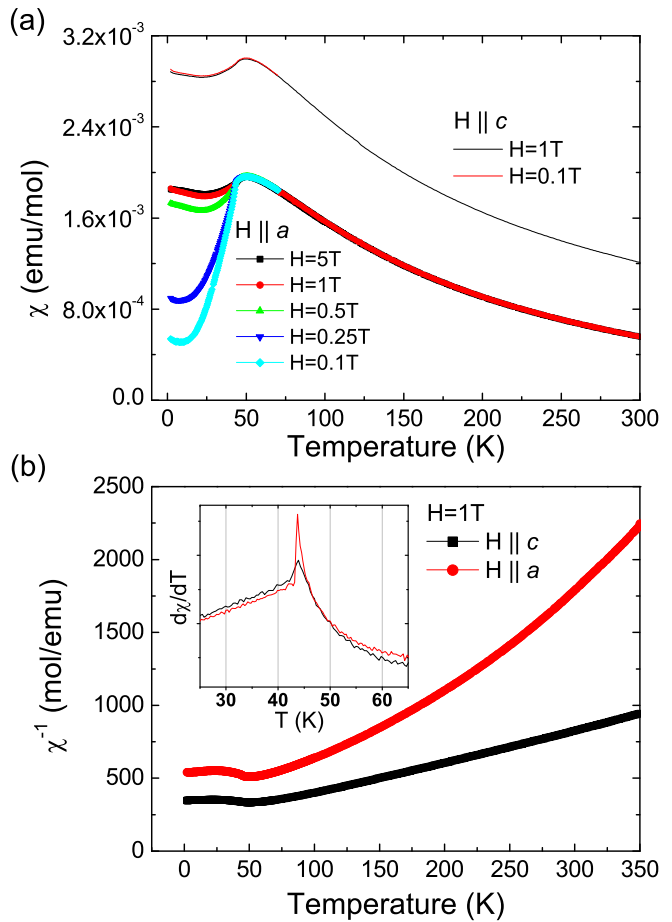}
\caption{(Color online) (a) The temperature dependence of magnetic
susceptibility, $\chi$, measured for various magnitudes and
directions of the magnetic field. (b) The corresponding temperature
dependence of $1/\chi$. The derivative (d$\chi$/dT) is shown in the
inset of panel (b). \label{figB}}
\end{figure}

\begin{figure}
\includegraphics[width=1\columnwidth]{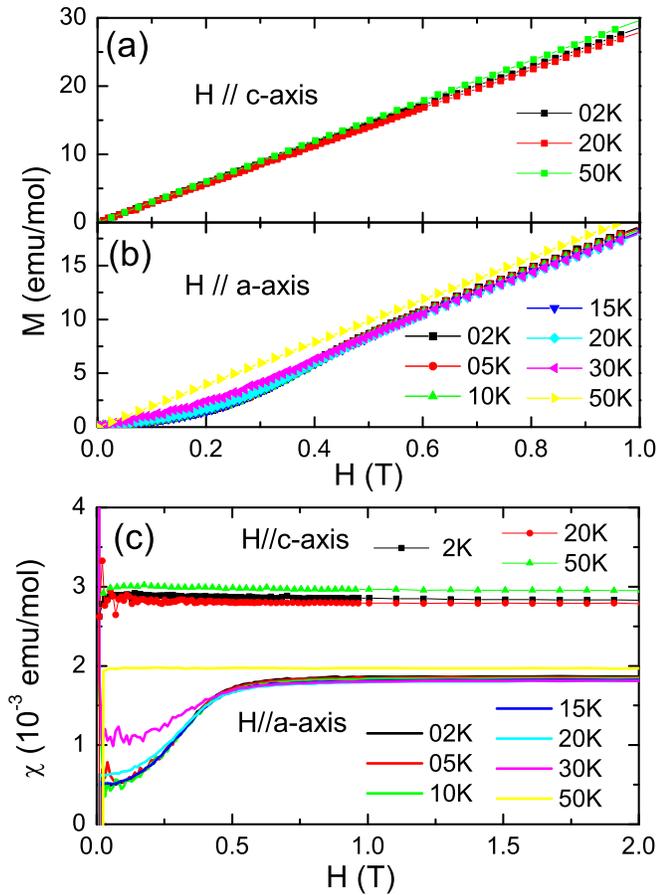}
\caption{(Color online) The field dependence of magnetization and
magnetic susceptibility with external magnetic field applied along
the $c$ and $a$ axes. \label{figC}}
\end{figure}

As shown in Fig.~\ref{figB}(a), $\chi$(T) increases on cooling,
reaches its maximum within a broad hump around $\sim$50~K, drops
abruptly around $\sim$43.5~K and decreases further until an upturn
below $\sim$20~K occurs. The broad hump in $\chi$(T) is indicative
for frustration or short-range fluctuations - a typical behavior
observed in low dimensional antiferromagnetic systems. The drop
around $\sim$43.5~K indicates the emergence of long-range AFM order
(i.e. \tn); see also d$\chi$/dT(T) in the inset of Fig.~2(b) which
exhibits a sharp peak at \tn$\sim$43.5~K. A strong anisotropy of
$\chi$(T) can be even observed at high temperature. We plotted the
temperature dependent inverse susceptibility ($\chi^{-1}$) in
Fig.~2b. A strong deviation from linear behavior can be observed
especially for H~$\|$~a even at high temperatures above \tn.
Apparently, a simple Curie-Weiss law is not applicable to \bco.
Note, that also ESR experiments \cite{q} report an anisotropic g
factor for \bco.

Whereas the susceptibility of \bco\ is field independent above \tn\
it exhibits an unusual field dependence below \tn\ (for H~$\|$~a) -
up to $\sim$1~T $\chi$ is increasing with increasing size of H.
Moreover, we measured the field-dependent magnetization and
susceptibility at different temperatures. As shown in Fig.~3(a,b)
the M-H curve is non-linear below 0.5-1~T for H~$\|$~a and for
temperatures below \tn. This behaviour might be indicative for a
metamagnetic transition with a kind of spin-reorientation. In
contrast to that we observe only linear M-H relations and
field-independent susceptibilities for H~$\bot$~a. Our results for
\bco\ are consistent with older studies in literature \cite{k,t}.

\subsection{Polarized neutron scattering}
\begin{figure}
\includegraphics[width=1\columnwidth]{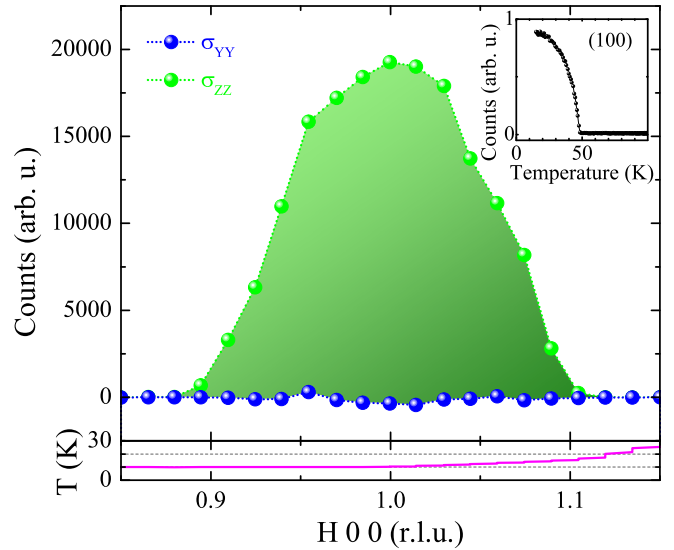}
\caption{(Color online) Polarization dependence of the magnetic
(100) reflection measured at the IN12 spectrometer. The green and
blue data points denote the magnetic neutron scattering intensities
in the $\sigma_{zz}$ and $\sigma_{yy}$ spin-flip channels with the
x-axis being parallel to the scattering vector and z-axis being
perpendicular to the scattering plane. The lower figure shows the
sample temperature at each measuring point of that scan which is
always well below T$_N$. The temperature dependence of the magnetic
(100) reflection is shown in the inset.  \label{figN}}
\end{figure}

\begin{figure}
\includegraphics[width=1\columnwidth]{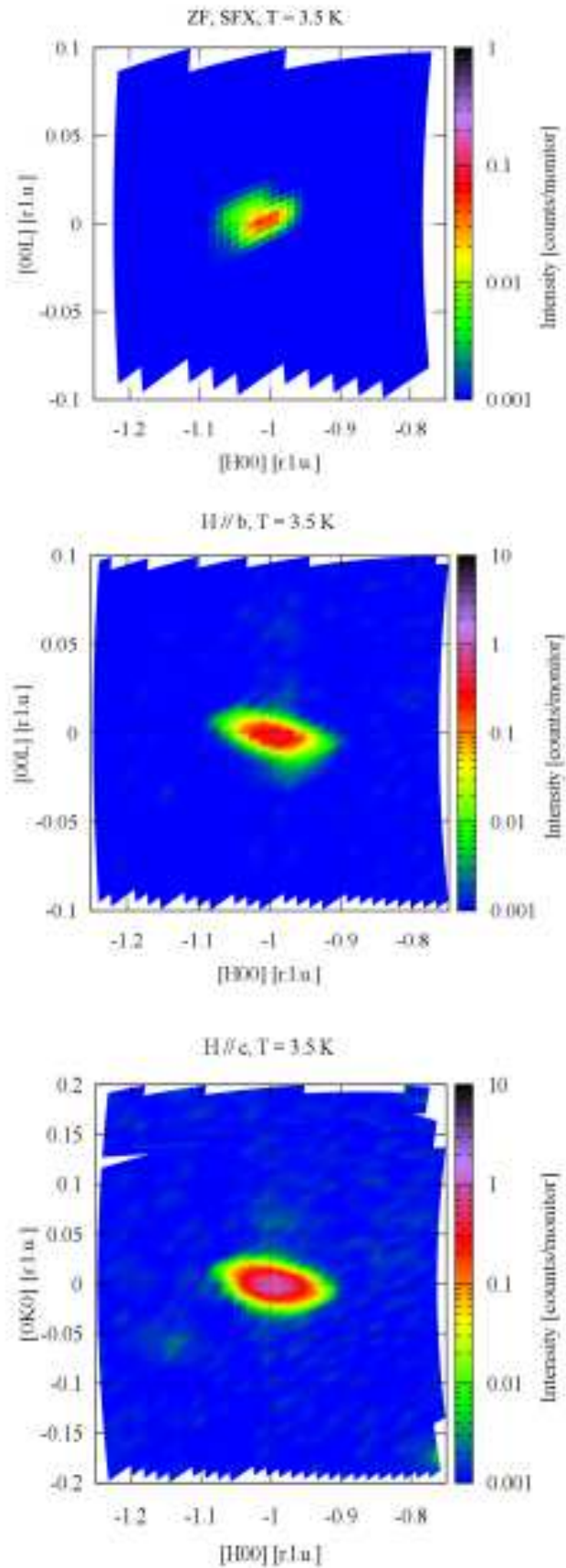}
\caption{(Color online) Intensity of the magnetic (100) reflection
measured at the DNS spectrometer in zero field (upper plot) and in
the magnetic fields $H$ of about 2.5~kG.  \label{figNb}}
\end{figure}

To resolve the long-standing controversy regarding the orientation
of ordered Cu spins\cite{g,j,k,l,m,n,o,p,q,r,s,t}, we studied the
magnetic structure of \bco\ by polarized neutron diffraction.
Figure~\ref{figN} shows the basically magnetic neutron scattering
intensities in the $\sigma_{zz}$ and $\sigma_{yy}$ spin-flip
channels. No intensity in the $\sigma_{yy}$ channel is observed.
Hence, the magnetic moments in \bco\ have no projection on the $c$
axis and are entirely aligned within the $ab$-plane. Thus, the
$c$-axis alignment of the copper spins as suggested in
Refs.~\cite{g,j} can be discarded and the alignment within the
$ab$-plane \cite{k} is corroborated. We obtained the same result in
a polarized neutron measurement at 3.5~K using the DNS spectrometer
at the FRM II, see Fig.~\ref{figNb}.

Magnetoelectrically induced ferroelectricity  (see
Sec.~\ref{sec:ferroelectricity}) and competing exchange interactions
\cite{x} prompted us to look for signs of incommensurate spin
ordering in \bco. In fields up to $\sim 0.25$~T we could not observe
any incommensurability  within the experimental resolution of our
measurements at the DNS spectrometer (see Fig.~\ref{figNb}). Further
high-resolution experiments in higher magnetic fields would be
desirable.

\subsection{Lattice parameters}

\begin{figure}
\includegraphics[width=1\columnwidth]{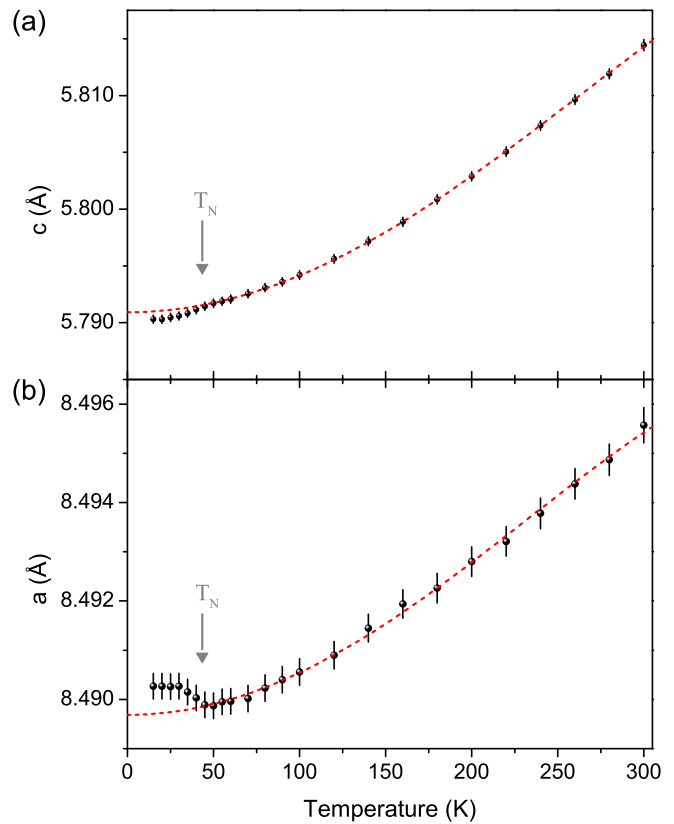}
\caption{(Color online) Temperature dependence of the lattice
parameters. The red dashed line indicates a fit to the data above
\tn. Below \tn\ an anomalous increase (decrease) of the $a$ ($c$)
lattice constant is observable. This observation is in agreement
with high-resolution thermal expansion measurements. \label{figX}}
\end{figure}

We have measured \bco\ by means of powder x-ray diffraction on a
powder x-ray diffractometer (Cu-K$\alpha$ radiation) that we
optimized for high resolution. Within these measurements we were not
able to find any indications for the occurrence of peak splittings
or superstructure reflections below 300~K. However, our high
resolution measurements are able to resolve an anomalous increase of
the $a$- and an anomalous decrease of the $c$-lattice constant on
cooling through \tn, see Fig.~\ref{figX}. This is indicative for
magneto elastic coupling as was also observed in other transition
metal compounds \cite{magcoup,magcoupB}. Our observed anomalous
increase of the lattice constants is in agreement with reported
thermal expansion measurements \cite{p}.

\subsection{Heat capacity}

\begin{figure}
\includegraphics[width=1\columnwidth]{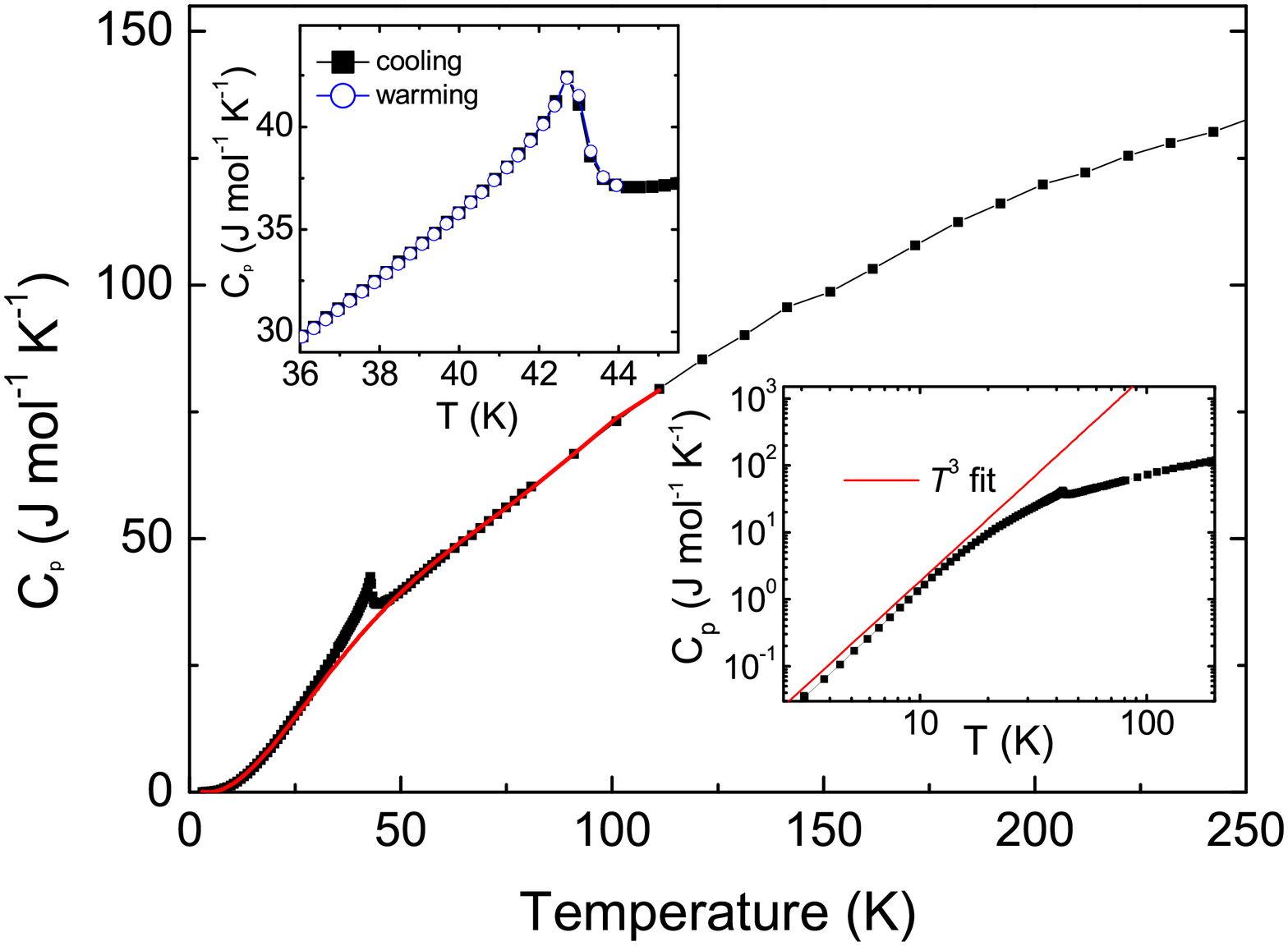}
\caption{(Color online) Temperature dependence of the specific heat,
$C_p$(T), for \bco\ single crystal. The red line represents the
background to be subtracted to estimate the change in entropy from
the anomaly peak around T$_N$. The upper right inset focuses on
$C_p$(T) in the region around the magnetic transition, and two sets
of $C_p$ data are shown, measured during warming (solid square) and
cooling processes (open circles) respectively. The $C_p$(T) is also
plotted on the log-log scale in the bottom right inset, with a T$^3$
(Debye) fit (red line) to the low temperature part of $C_p$(T).
\label{figD}}
\end{figure}

Further evidence for long-range antiferromagnetic order in \bco\ is
revealed from our specific heat (\emph{C}$_p$) measurements. As
shown in Fig.~\ref{figD},  \emph{C}$_p$(T) shows a $\lambda$-shaped
anomaly at \tn. We have measured the specific heat also on heating
and on cooling around \tn, see the upper left inset of
Fig.~\ref{figD}. The absence of any hysteresis indicates a second
order phase transition. This is also consistent with the absence of
any structural transition at \tn\ observed in our powder x-ray
diffraction measurements.

Both, lattice vibrations (phonons) and magnetic excitations
(magnons) contribute to the specific heat $C_p$. At low temperatures
the phonon contribution of $C_p$ obeys the Debye T$^3$ power law. As
can be seen in the lower right inset in Fig.~\ref{figD}, the low
temperature part of $C_p$(T) can be fitted quite well with a
function proportional to T$^3$, thus, suggesting that also the
magnetic contribution of $C_p$ obeys a T$^3$ law. The T$^3$
dependence  - rather than T$^{3/2}$ behaviour - further confirms
long-range AFM ordering in \bco. However, it remains difficult to
extract the pure magnetic part of $C_p$(T) accurately due to the
absence of nonmagnetic isostructural materials in order to estimate
the contribution of the lattice vibrations to the total specific
heat. To get a rough estimate of the entropy removed by the magnetic
ordering, which is remarkably indicated by the peak of Cp(T) around
\tn, we use a polynomial fit to the high-temperature ($>$60~K) and
low-temperature ($<$25~K) regimes for the background (the red solid
line shown in main panel of Fig.~\ref{figD}). The entropy obtained
from integrating the peak in C$_p$(T)/T (for 25~K~$<$~T~$<$~60~K)
amounts to 1.82~J~mol$^{-1}$~K$^{-1}$ which is only 32\%\  of the
full magnetic entropy associated with R$\cdot$ln(2S+1)
$\thickapprox$~5.76~J~mol$^{-1}$~K$^{-1}$ as expected for an
S~$=$~1/2 spin system. Most of the remaining magnetic entropy is
considered to be removed by the short-range ordering well above \tn,
as usually observed in the low-dimensional frustrated spin systems.

\subsection{Anisotropic dielectric properties as zero field}

In view of the tetragonal symmetry of \bco\ we measured the
dielectric constant for an electric field $E$ applied either within
the $ab$ plane, \ep$_{ab}$, or along the $c$-axis, \ep$_c$. Two thin
plate samples that were cut from our floating-zone-grown single
crystal were used to investigate the anisotropic dielectric
properties.

\begin{figure}
\includegraphics[width=1\columnwidth]{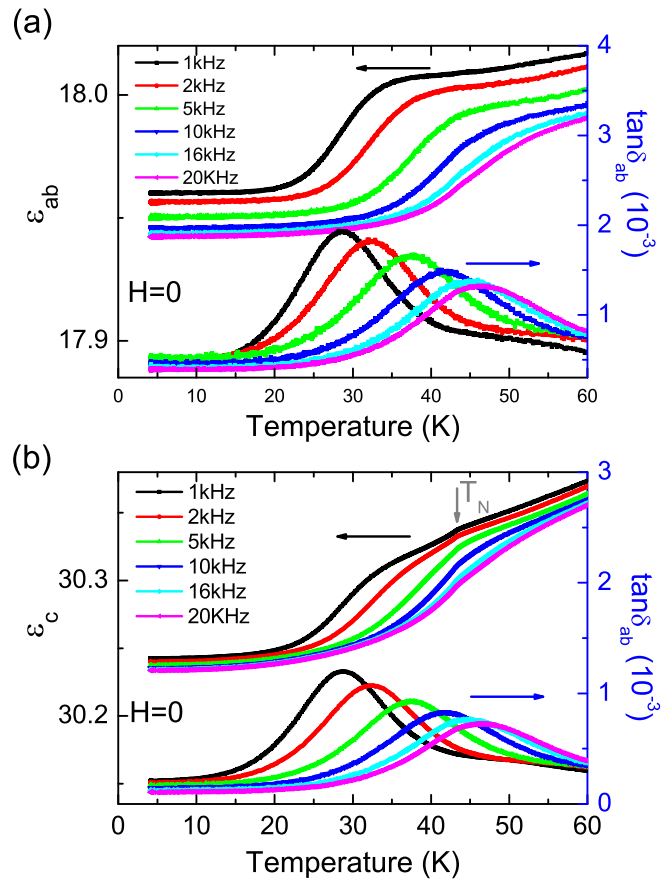}
\caption{(Color online) Temperature dependence of dielectric
contants: (a)  \ep$_{ab}$ (E $\bot$ $c$-axis) and (b) \ep$_c$ (E
$\|$ $c$-axis). Also shown is the corresponding dielectric loss
(tan$\delta$). All data has been measured in zero field (H~=~0). The
vertical arrow in (b) marks \tn\ where the slight kink occurs.
\label{figE}}
\end{figure}

Fig.~\ref{figE} shows the temperature-dependent \ep$_c$ and
\ep$_{ab}$ measured in zero field for different frequencies. We
observed a strong anisotropy in permittivity:  \ep$_{ab} \sim 18$ is
much lower than \ep$_{c} \sim 30$, which resembles the anisotropy in
magnetic susceptibility discussed above.

Both \ep$_{c}$ and \ep$_{ab}$ slightly decrease with decreasing
temperature and saturate at lowest temperature. \ep$_{ab}$ and
\ep$_{c}$ also exhibit a strong frequency dependence characteristic
of a Debye-type dipolar relaxation-like behavior. For analysis of
the frequency dispersion, the corresponding dielectric loss
(tan$\delta$) curves are also shown. At each measuring frequency the
loss exhibits a peak corresponding to the knee-like change of \ep.
As frequency increases, the peak shifts to higher temperatures with
a concomitant decrease of magnitude. The peak temperature and the
frequency can be fitted with the Arrhenius law, $f = f_0 \exp(- Q /
k T_{peak})$, where $f$ is the characteristic frequency, $k$ is the
Boltzmann constant, T$_{peak}$ is the peak temperature, $f_0$ is a
pre-factor and $Q$ is the activation energy. Fitting the two sets of
data from Fig.~\ref{figE}(a) and (b) yields nearly the same
activation energy of $\sim$22~meV. We note that in Raman scattering
experiments on \bco\ a featured mode around 150~cm$^{-1}$ was
observed at low temperature, which results from a two-magnon
scattering process. The corresponding energy ($\sim 20$~meV) is
close to $Q$ derived from dielectric dispersion analysis, which
suggests that the observed dielectric relaxation at low temperatures
can result from the coupling of electric dipoles to magnons.

Across the magnetic ordering temperature \tn, a weak `kink' appears
in  \ep$_{c}$(T) at all frequencies. However,  no corresponding
anomalous behavior was found in the dielectric loss indicating the
absence  of ferroelectric transition in zero field. This conclusion
is corraborated by our pyroelectric measurements discussed below. No
anomaly is observed in  \ep$_{ab}$(T).

\subsection{In-plane dielectric response in an applied magnetic field}

Next we studied the in-plane dielectric properties under applied
magnetic fields. We present the data measured at 1~kHz because of
the relatively large separation of the dispersion peak from \tn\ and
a better signal-to-noise ratio at low frequencies.

\begin{figure}
\includegraphics[width=1\columnwidth]{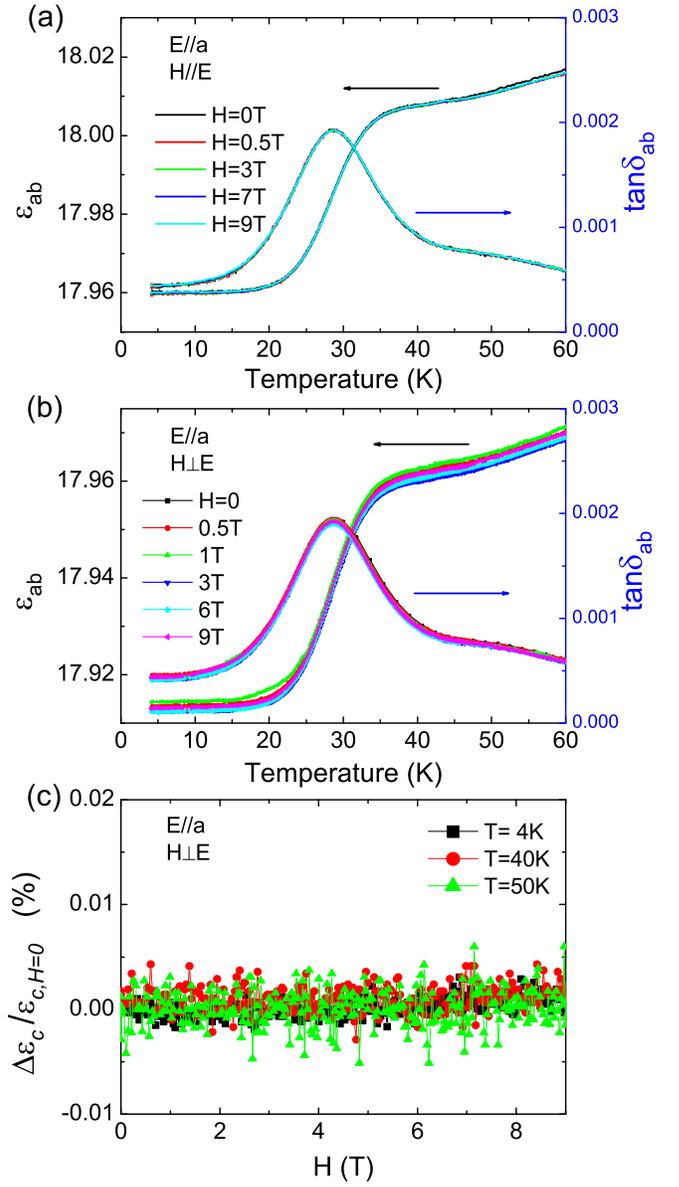}
\caption{(Color online) Temperature dependence of in-plane
dielectric constants \ep$_{ab}$ and corresponding loss for (a)
H~$\|$~E and (b) H~$\bot$~E. The corresponding dielectric loss
(tan$\delta$) is also shown. The measurement frequency amounts to
1~kHz. (c) The magnetic field induced change of the dielectric
constant (measured at 1~kHz) at different temperatures (4~K, 40~K
and 50~K).    \label{figF}}
\end{figure}

We applied $H$ parallel and perpendicular to the measuring electric
field $E$. Up to 9~T no change can be observed in \ep$_{ab}$  and
tan$\delta_{ab}$, as shown in Fig.~\ref{figF}. The corresponding
pyroelectric measurements also show no electric polarization (not
shown). We also investigated the in-plane magnetodielectric effect
at temperatures below and above \tn\ for several values of the
applied field $H$. As shown in Fig.~\ref{figF}(c) for $H \perp E$ no
magnetodielectric effect is observed. Similar results were found for
$H \parallel E$.

\subsection{Out-of-plane dielectric properties in an applied field and magnetically-induced ferroelectricity}
\label{sec:ferroelectricity}

In this section we describe the effect of magnetic field on the
dielectric permitivity  along the tetragonal $c$ axis,  \ep$_{c}$.
We show that under an applied magnetic field \bco\ becomes
ferroelectric with the field-dependent electric polarization, $P_c$,
parallel to the $c$ axis.

\subsubsection{Out-of-plane dielectric response in $H \parallel c$}

\begin{figure}
\includegraphics[width=1\columnwidth]{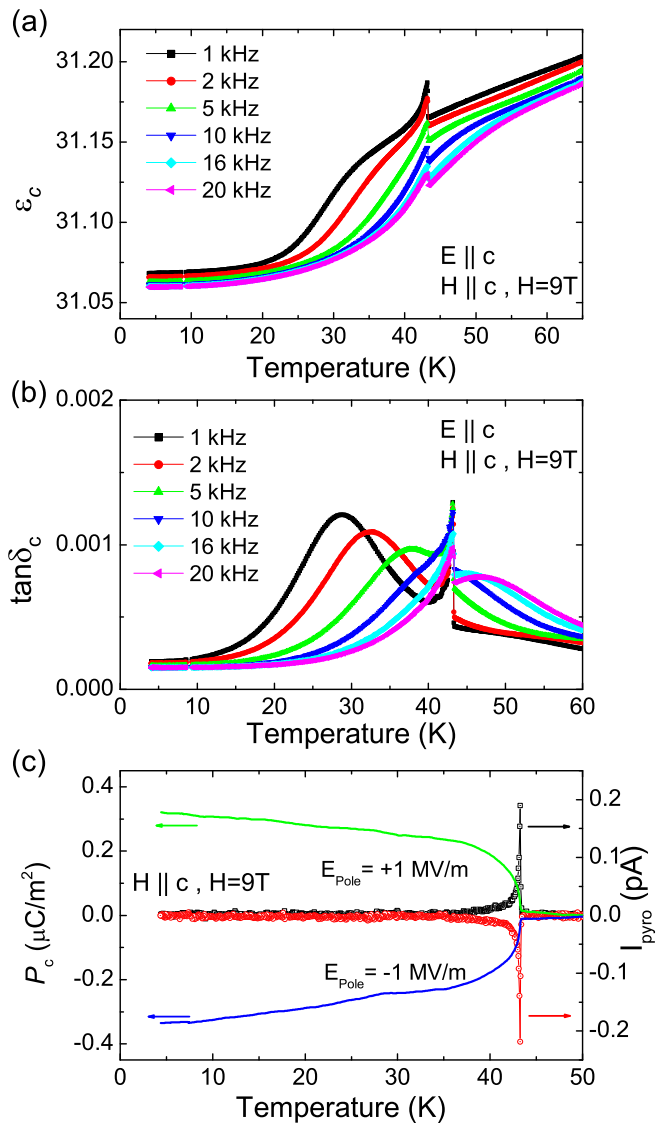}
\caption{(Color online) Temperature dependence of  (a) the c-axis
dielectric constant \ep$_c$ and (b) the tan loss for different
frequencies measured in a field of H~=~9~T along the $c$-axis. (c)
the pyroelectric currents (black and red dots) measured with both
positive and negative poling electric fields of 1~MV/m applied. The
corresponding polarization (P$_c$ , green and blue solid lines)
obtained by integrating the pyroelectric currents from above \tn.
\label{figG}}
\end{figure}

We performed dielectric measurements for a variety of magnetic
fields applied along the $c$-direction. Figure~\ref{figG} shows the
detailed results for $H$~=~9~T. The most remarkable observation is
the sharp $\lambda$-shaped peak observed in both \ep$_{c}$(T) and
tan$\delta_c$ for all frequencies, which is indicative of a
ferroelectric transition. The peak position is independent of
frequency and coincides with \tn\ determined from magnetic
susceptibility and specific heat measurements. Compared with the
zero field data (Fig.~\ref{figE}(b)) shows that the
frequency-dependent dielectric relaxation behavior is essentially
unaffected by the external magnetic field.

The direct evidence for the ferroelectric nature of the transition
at \tn\ is given by our pyroelectric current measurements shown in
Fig.~\ref{figG}(c). A small but very sharp peak emerges as
temperature approaches \tn. The pyroelectric current drops to zero
above \tn. The corresponding $c$-axis electric polarization, $P_c$,
obtained by integrating the pyroelectric current, appears below \tn\
and reaches its (rather small) saturation value of about
0.18$\mu$C/m$^{2}$. The reversal of the poling electric field
changes sign of $P_c$. This proves the emergence of ferroelectricity
in \bco under the applied magnetic field.

\begin{figure*}
\includegraphics[width=1.5\columnwidth]{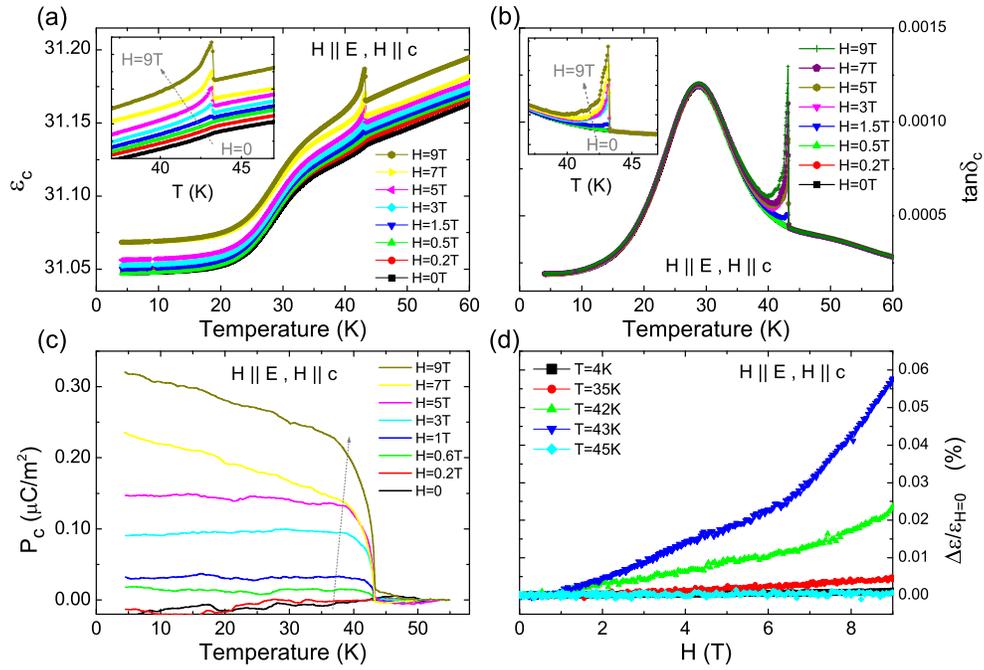}
\caption{(Color online) Temperature dependence of (a) \ep$_c$, (b)
tan loss and (c) polarization measured in different fields along the
$c$ axis. (d) The magnetic field induced change of \ep$_c$ at
different temperatures (from 4~K to 45~K). The measuring frequency
amounts to 1~kHz.   \label{figH}}
\end{figure*}

Figure~\ref{figH} shows \ep$_c$(T) measured at the frequency 1~kHz
for different applied magnetic fields ranging from 0 to 9~T.
For magnetic fields below $\sim$$\sim 0.5$~T the small kink around
\tn\ observed in zero field remains unchanged.
As the magnetic field is increased further, a dielectric peak
appears (see e.g.  the blue curve measured at 1.5~T in the inset of
Fig.~\ref{figH}(a)).
This dielectric peak develops gradually with increasing $H$ and no
saturation is observed up to our instrumental limit of 9~T.
The transition temperature  \tn\ at which the peak is observed is
almost unaffected by $H$.
The same behavior is observed in the corresponding loss curves shown
in Fig.~\ref{figH}(b).
Hence, the dielectric anomaly associated with the onset of
ferroelectricity in \bco\ appears for applied magnetic fields
($>$$\sim 0.5$~T) and becomes more pronounced as the magnetic field
increases.
The results for field and temperature dependence of $P_c$
(Fig.~\ref{figH}(c)) clearly show that \bco\ becomes ferroelectric
under an applied magnetic field $H \| c$.

We also studied the magnetodielectric coupling for $H \| c$, i.e.
the dependence of the dielectric constant on an applied magnetic
field. Figure~\ref{figH}(d) shows the relative change of the
permitivity, $\frac{\Delta \varepsilon}{\varepsilon} =
\frac{\varepsilon_c(H)-\varepsilon_c(0)}{\varepsilon_c(0)}$, for
several temperatures below and above the transition temperature,
\tn. A  prominent magnetodielectric effect is only observed for $T$
close to  \tn. At $T = 43$~K, the value of  $\frac{\Delta
\varepsilon}{\varepsilon}$ grows with increasing field $H$ ($>$1~T)
reaching $\sim 0.06$\%\ at 9~T.

\subsubsection{Out-of-plane dielectric response in $H \parallel a$}
\label{H||a}

\begin{figure*}
\includegraphics[width=1.5\columnwidth]{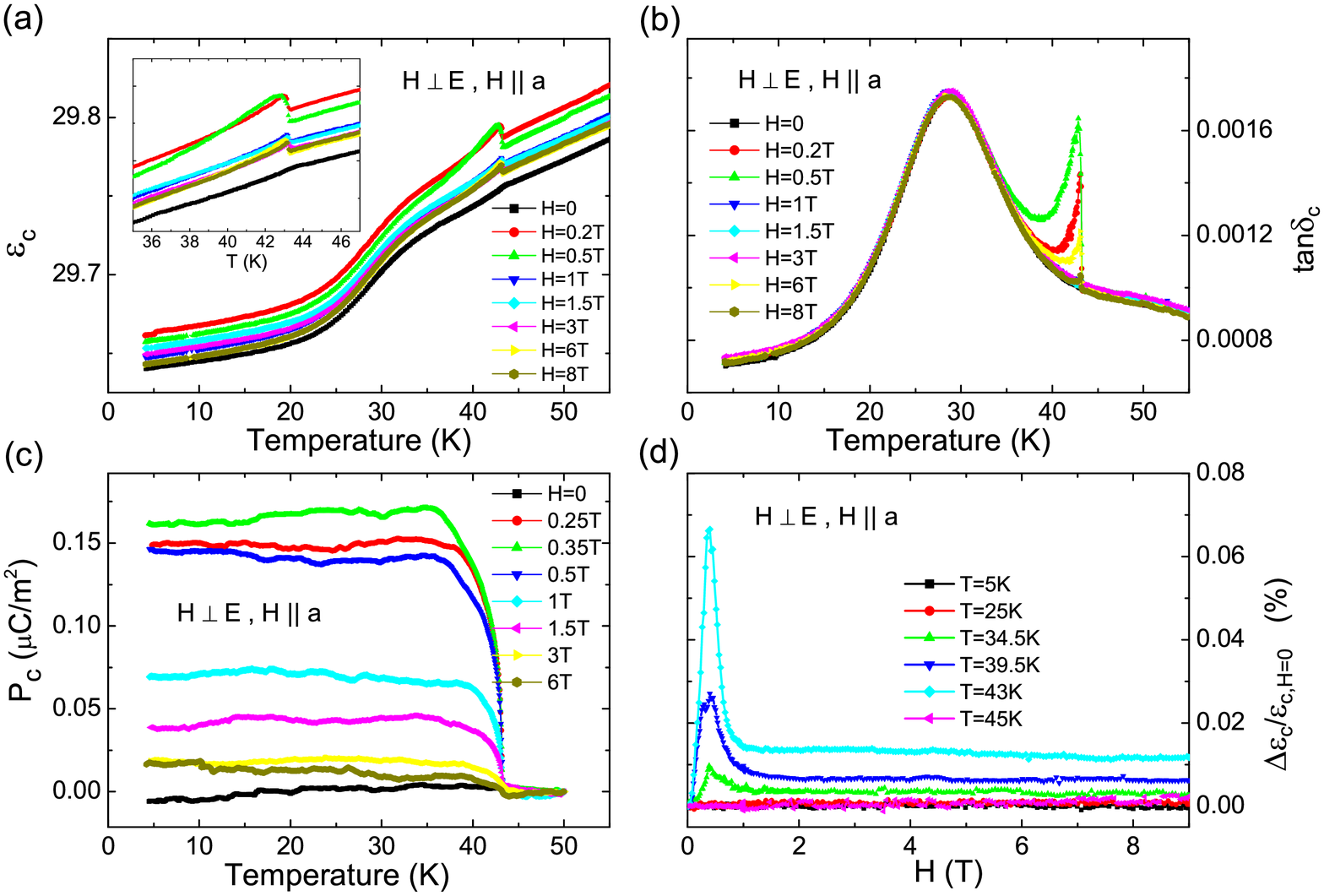}
\caption{(Color online) Temperature dependence of (a) \ep$_c$, (b)
tan loss and (c) the polarization measured in different magnetic
fields applied along the $a$ direction. (d) The magnetic field
induced change in \ep$_c$ at different temperatures (from 5~K to
45~K). The measuring frequency is 1~kHz.   \label{figI}}
\end{figure*}

We performed similar measurements for magnetic fields applied along
the $a$ direction ($H \perp E$). The results are shown in
Fig.~\ref{figI}.
In contrast to the behavior observed for $H \| c$, the dielectric
peak in \ep$_c$(T) at \tn\ is discernable already at very low
magnetic fields.
At about $\sim 0.5$~T the peak reaches its maximum size and becomes
smaller as $H$ is increased further (see the inset in
Fig.~\ref{figI}(a)).
Concomitantly with the peak in $\varepsilon_c$, a sharp peak in the
dielectric loss, tan~$\delta_c$, appears and disappears (see
Fig.~\ref{figI}(b)).

\begin{figure}[!h]
\includegraphics[width=1\columnwidth]{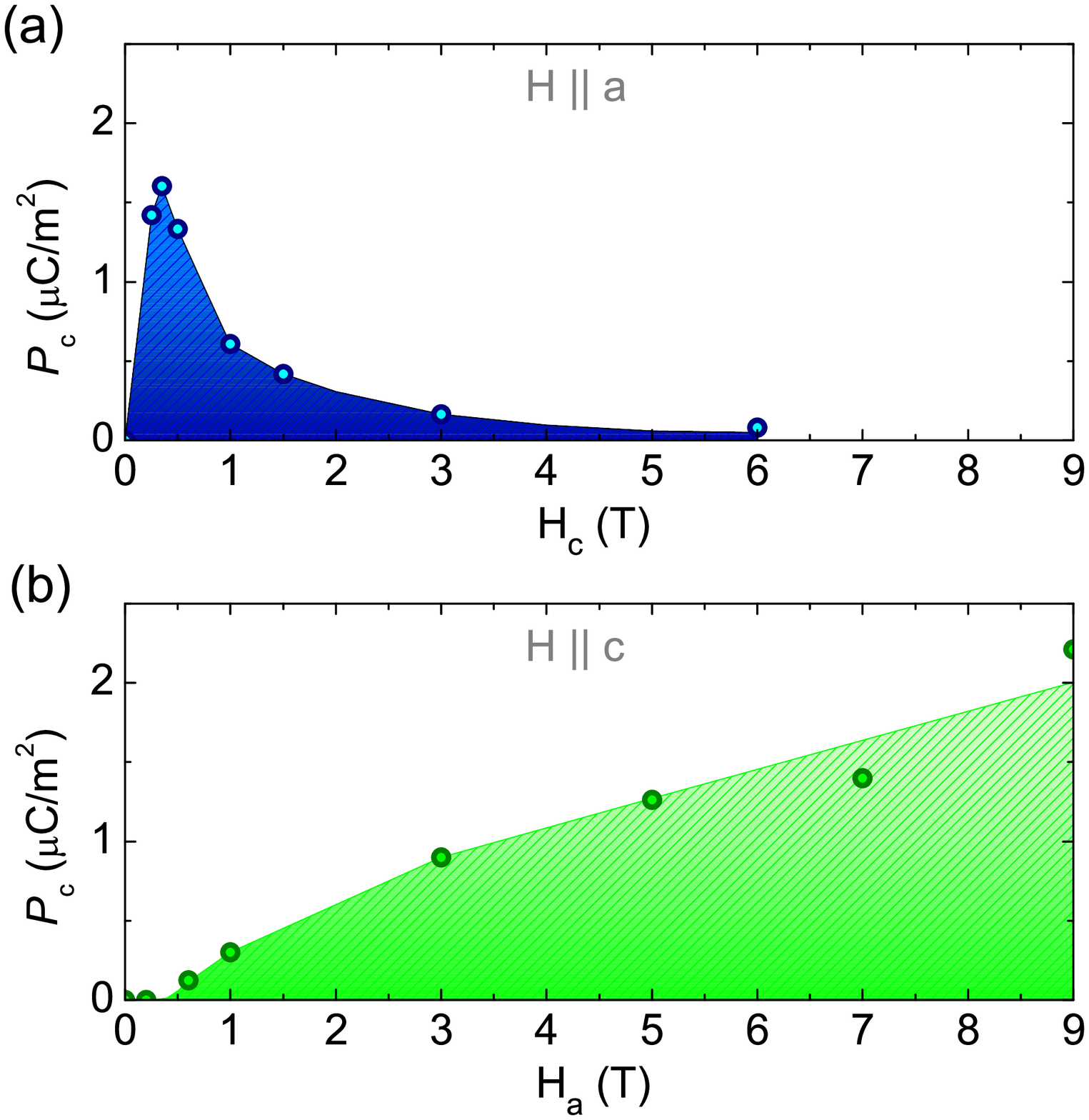}
\caption{(Color online) Phase diagram of Bi$_2$CuO$_4$ at 35~K which
contains the information about the magnetic field dependence of the
electric polarization P$_c$ for two different field directions: (a)
H~$\|$~a and (b) H~$\|$~c.    \label{figJ}}
\end{figure}

Fig.~\ref{figI}(c) shows the electric polarization $P_c$ measured in
$H \| a$. A low field of just 0.2~T induces ferroelectricity in
\bco\ with a spontaneous polarization of about 0.14$\mu$C/m$^2$ at
4~K. After $P_c$ reaches its maximum value, it decreases upon
further increase of the magnetic field and vanishes around 6~T.

Finally, we also studied the magnetodielectric effect for $H \| a$.
As for $H \| c$, there is no discernable magnetodielectric effect at
temperatures far from \tn. Only around the dielectric peak region we
are able to observe a prominent magnetodielectric effect in low
fields. At 43~K, $\frac{\Delta \varepsilon}{\varepsilon}$  rises
sharply with increasing field up to its maximum value at about $\sim
0.5$~T, and then decreases upon further increase of $H$.

\section{Discussion}

In zero field, previous neutron diffraction measurements on \bco\
reveal a k~=~0 collinear spin structure with parallel spins in the
chains running along the $c$  and antiparallel spins in neighboring
chains. However, the orientation  of ordered magnetic moments is
poorly understood and remains controversial. In other magnetic
Cu$^{2+}$ (S~=~1/2) compounds with the Cu\emph{L}$_4$ (\emph{L}=O,
Cl, Br) square-planar coordination (e.g. in CuCl$_2\cdot$2H$_2$O,
CuCl$_2$, CuBr$_2$ or LiCuVO$_4$ ) an   easy-plane anisotropy is
observed. Our polarized neutron data suggest that this is also the
case for \bco.

Our observation of a metamagnetic behavior in low in-plane fields
(of about $\sim 0.5$~T at 2~K) is consistent with previous findings
\cite{t,k}. The non-linearity of the $M(H)$ curve likely originates
from the field-dependent volume fraction of  the four
antiferromagnetic domains, in which spins are (anti)parallel to the
$a$ or $b$ axis. We cannot exclude, however, a more complex
spin-reorientation transition taking place at $H_a \sim 0.5$~T.

Concomitant with the emergence of a long range magnetic ordering at
\tn, we observe a transition into a ferroelectric state with the
polarization along the $c$-direction, which appears under an applied
magnetic field. This transition is manifest in both the dielectric
constant, \ep$_c(T,H)$, and polarization, $P_c(T,H)$,  measurements.
The transition occurs for magnetic fields applied either in the
$ab$-plane or along the $c$ direction. The electric polarization
$P_c$ as a function of the magnetic field along the $a$ and $c$ axes
at 35~K (at which the polarization nearly saturates) is shown in
Fig.~\ref{figJ}. As we discuss below, the magnetoelectric response
provides an important additional information on magnetic ordering in
\bco.

Although the magnetic field dependence of the electric polarization
is nonlinear, the most likely source of the field-induced
ferroelectricity is the linear magnetoelectric effect
\cite{xa,xb,xc}. We assume that the antiferromagnetic order
parameter in \bco\ is,
$\mathbf{L}=\mathbf{M}_1-\mathbf{M}_2+\mathbf{M}_3-\mathbf{M}_4$,
where $\mathbf{M}_i$, $(i=1,2,3,4)$ is the sublattice magnetization
labeled by the four independent Cu sites occupying the 4c Wykoff
positions in the crystallographic unit cell with the coordinates
$\mathbf{r}_1 = (1/4, 1/4, z)$,
 $\mathbf{r}_2 = (3/4, 3/4, 1/2-z)$, $\mathbf{r}_3 = (1/4, 1/4, 1/2+z)$ and $\mathbf{r}_4 = (3/4, 3/4,-z)$. Spatial inversion, $\mathbf{r} \rightarrow - \mathbf{r}$ interchanges the sites 1 and 4 and the sites 2 and 3, so that under inversion $\mathbf{L}$ changes sign.  The inversion symmetry breaking by this antiferromagnetic ordering allows for a linear magnetoelectric effect. We note that the non-centrosymmetric antiferromagnetic order is incompatible with the interpretation of the metamagnetic behavior in terms of a weak ferromagnetic moment \cite{yy}.

The free energy density describing the linear mangetoelectric
coupling contains 4 terms allowed by \emph{P4/ncc} symmetry of the
crystal lattice,
 \begin{eqnarray}
 f_{\rm me} = &-& g_1 (L_a H_a + L_b H_b) E_c - g_2 (H_aE_a + H_b E_b) L_c\nonumber \\
 &-&  g_3 (E_a L_a + E_b L_b) H_c - g_4 L_c H_c E_c,
 \end{eqnarray}
where $g_\alpha$ $(\alpha = 1,2,3,4)$ are the coupling constants.
The $c$-component of the electric polarization induced by an applied
magnetic field is then
\begin{equation}
P_c =-\partial f_{\rm me} / \partial E_c =  g_1 (L_a H_a + L_b H_b)
+ g_4 L_c H_c. \label{eq:Pz}
\end{equation}
Similar expressions can be obtained for $P_a$ and $P_b$, but since
no significant in-plane electric polarization was observed, the
coupling constants $g_2$ and $g_3$ must be  relatively small.

 We first discuss the magnetoelectric behavior of \bco\ in H~$\|$~a (see Figs.~\ref{figI}(c) and \ref{figJ}(a)). Our polarized neutron scattering data show that in zero field the antiferromagnetic vector  $\mathbf{L}$ lies in the $ab$ plane.  It can be oriented either along the crystallographic axes  (the $\pm[100]$ and $\pm[010]$ directions) or along the diagonals (the $\pm[110]$ and $\pm[1\bar{1}0]$ directions). In the latter case, the magnetic field parallel to the $a$ axis would result in a gradual rotation of  $\mathbf{L}$  toward the $[100]$ plane.  Instead, we observe a metamagnetic  transition at $H_a \sim 0.5$ T (see Fig.~\ref{figC}), which indicates that in zero field \bco\  is divided into domains with $L \| a$ and $L \| b$. The zero-field magnetic susceptibility is then the average of the longitudinal ($L \parallel H$) and transverse ($L \perp H$) magnetic susceptibilities. Above the metamagnetic transition the $L_a$-domains disappear and the magnetic susceptibility becomes purely transversal, which explains why above $\sim 0.5$~T it increases approximately by factor of $2$  (see  Fig.~\ref{figC}). The peculiar dependence of $P_c$ on $H_a$ shown in Fig. ~\ref{figJ}(a) is consistent with this scenario.  As follows from Eq.(\ref{eq:Pz}), the linear magnetoelectric coefficient $\alpha_{ca} = \frac{\partial P_c}{\partial H_a} = g_1 L_a$, which is why $P_c$ grows linearly with the field below $\sim 0.5$~T, when the $L_a$-domains are present, and decreases above $\sim 0.5$~T, when these domains disappear.

Below we discuss the spin-reorientation transition in more detail.
Because of tetragonal symmetry of \bco, the $L \| b$ state has a
lower free energy than the $L \| a$ for an arbitrarily small $H_a$.
However, in low applied magnetic fields the motion of domain walls
separating the $L \| a$ and $L \| b$ domains can be hindered by
pinning and the ferroelectric state with $L_a \neq 0$ can survive as
a metastable state. There is, however, an upper critical field,
$H_{\rm cr}$, above which the ferroelectric state becomes locally
unstable and disappears. This can explain the nature of the
metamagnetic transition observed at $\sim 0.5$~T.

This transition as well as the observed magnetoelectric response can
be described using the phenomenological Landau theory. Assuming that
both the vector order parameter and the applied magnetic field are
confined to the $ab$ plane, while the electric field is parallel to
the $c$ axis, we can write the free energy density of \bco\ in the
form,
\begin{eqnarray}
&~& f \approx \frac{a}{2} L^2 + \frac{b_1}{4} L^4 + \frac{b_2}{2}
L_a^2 L_b^2
+ \frac{c_1}{2}L^2 H^2 + \frac{c_2}{2} (\mathbf{L}\cdot\mathbf{H})^2 \nonumber \\
&~& + c_3 L_a L_b H_a H_b
 - \frac{\chi_0}{2} H^2 - g_1 (\mathbf{L}\cdot\mathbf{H}) E_c,
 \label{eq:Landau}
\end{eqnarray}
where $L^2 = L_a^2 + L_b^2$, $H^2 = H_a^2 + H_b^2$,
$(\mathbf{L}\cdot\mathbf{H}) = L_a H_a + L_b H_b$, $\chi_0$ is the
background magnetic susceptibilty and $a = \alpha (T - T_N(0))$,
$T_N(0)$ being the transition temperature in zero magnetic field.
The quartic anisotropy coefficient, $b_2 > 0$, favors $\mathbf{L}$
(anti)parallel to the $a$ and $b$ axes, while $c_2 > 0$ favors
$\mathbf{L} \perp \mathbf{H}$.

Minimizing the free energy Eq.(\ref{eq:Landau}) with respect to
$L_a$ for $L_b = H_b = 0$, we obtain the order parameter in the
metastable state,
\begin{equation}
L_a^2 = \frac{\alpha(T_N(H_a) - T)}{b_1}, \label{eq:L_a}
\end{equation}
where $T_N(H_a) = T_N(0) - \frac{(c_1+c_2)}{\alpha} H_a^2$ is the
transition temperature in the applied field. The
(temperature-dependent) critical field,  $H_{\rm cr}$, above which
this state is unstable, is given by
\begin{equation}
H_{\rm cr}^2 = \frac{\alpha(T_N(0) - T) b_2}{b_1 c_2 + b_2(c_1 +
c_2)}.
\end{equation}

The electric polarization, given by Eq.(\ref{eq:Pz}), is
approximately propoportional to $H_a$, for $H_a < H_{\rm cr}$ (the
magnetic field dependence of $L_a$ (see Eq.(\ref{eq:L_a})) is weak,
since our experiment shows that the transition temperature is
practically field independent). Finally, the dielectric
susceptibility associated with the antiferromagnetic order is given
by
\begin{equation}
\chi_c = \left.\frac{\partial P_c }{\partial E_c} \right|_{E_c = 0}
= \chi_L g_1^2 H_a^2, \label{eq:chic}
\end{equation}
where $\chi_L$ is the antiferromagnetic susceptibility: $\chi_L =
\left(\frac{\partial^2 f}{\partial L_a^2}\right)^{-1} \propto |T -
T_N(H_a)|^{-1}$ near the transition (see Ref.~[\onlinecite{ya}] ,
where also the effect of spin fluctuations on  the critical behavior
of the dielectric susceptibility of a linear magnetoelectric
material is discussed). The dielectric susceptibility diverges
because  in the applied magnetic field a linear magnetoelectric
material becomes ferroelectric, since for $H_a \neq 0$ the
antiferromagnetic order, $L_a$  is linearly coupled to the electric
field, $E_c$. Equation (\ref{eq:chic}) is consistent with our
observations (see Sec.~\ref{H||a}): the magnetic field dependence of
$\varepsilon_c$ is only strong close to $T_N$, where the divergence
of $\chi_c$ compensates for the weakness of the magnetoelectric
response.

 The explanation of the magnetoelectric response in H~$\|$~c (see Figs.~\ref{figH}(c) and \ref{figJ}(b)) is less straightforward. According to Eq.(\ref{eq:Pz}),   the linear magnetoelectric coefficient $\alpha_{cc} = \frac{\partial P_c}{\partial H_c} = g_4 L_c$ is only nonzero if $L_c \neq 0$ . However, $L_c$ is zero in zero field and H~$\|$~c is unlikely to result in a transition that rotates $\mathbf{L}$ out of the $ab$ plane.  Also the linear dependence of $M_c$ on $H_c$ (Figs.~\ref{figC}(a) and (c)) shows the absence of any transition. One possible explanation is a small in-plane magnetic field component resulting from a small misalignment of the sample or demagnetizing fields.  Note that $\alpha_{cc}$ is $\sim 15$ times smaller than $\alpha_{ca}$ for $H_a < 0.5$ T.

Another possible explanation is  that the actual symmetry of \bco\
is lower than tetragonal. We did not, however, find  any indication
for a symmetry lowering, even though  our high-resolution powder
X-ray diffraction measurements allow us to resolve tiny thermal
expansion effects (see Fig.~\ref{figX}). Ferroelectricity in  \bco\
can, in principle, be induced by a complex magnetic ordering, such
as  an incommensurate spin spiral with a small wave vector. The
rotation of the spiral plane in an applied magnetic field, which
transforms a non-ferroelectric helical spiral into a ferroelectric
cycloidal spiral,  can mimic linear magnetoelectric response, as was
observed in the CoCr$_2$Se$_4$ spinel \cite{yb,yc}. However, also in
the spiral state scenario it is difficult to explain why the
out-of-plane electric polarization is induced by both in-plane and
out-of-plane magnetic fields.

\par
\section{Conclusions}
A combination of several experimental  techniques was applied to
clarify the nature of magnetic ordering in  floating zone grown
\bco\ single crystals.
Polarized neutron scattering  measurements clearly show that ordered
magnetic moments of Cu ions are confined to the $ab$ plane, which
resolves the long standing controversy.
In addition, we observe a ferroelectric polarization parallel to the
$c$ axis induced by applied magnetic fields, which is consistent
with inversion symmetry breaking by the C-type antiferromagnetic
ordering.
We argue that the electric polarization in the magnetic field
parallel to the $a$ axis originates from the linear magnetoelectric
effect in a metastable state, which disappears above a  critical
magnetic field.
The relatively weak response observed in the magnetic field parallel
to the $c$ axis does not seem to be compatible with symmetry of
\bco\ and remains a puzzle.
Further studies of the magnetic states of this material are
necessary.

\begin{acknowledgments}
The authors would like to thank L.~H.~Tjeng and O.~Stockert for
fruitful discussions and Ch. Becker, T. Mende and S. Wirth for their
support on the experimental construction of the dielectric
measurement system.
\end{acknowledgments}

\end{document}